\documentclass[runningheads]{llncs}
\usepackage{graphicx}
\usepackage{multirow}                %multirow for format of table
\usepackage{bm}
\usepackage{amsmath}
\usepackage{amssymb}
\usepackage{makecell}
\usepackage[misc]{ifsym}
\usepackage{url}
\usepackage{array}
\usepackage{cite}

\usepackage{graphicx}
\usepackage{subfigure}
\usepackage{multirow}
\usepackage{makecell}
\usepackage{color}
\usepackage{cite}
\usepackage{caption}
\newcommand{\PreserveBackslash}[1]{\let\temp=\\#1\let\\=\temp}
\newcolumntype{C}[1]{>{\PreserveBackslash\centering}p{#1}}
\newcolumntype{R}[1]{>{\PreserveBackslash\raggedleft}p{#1}}
\newcolumntype{L}[1]{>{\PreserveBackslash\raggedright}p{#1}}
\usepackage[colorlinks,
            urlcolor=red,
            hyperfootnotes = True,
            citecolor=green]{hyperref}

\def\mP{\mathcal{P}}

\begin{document}
\title{InDuDoNet: An Interpretable Dual Domain Network for CT Metal Artifact Reduction}
\titlerunning{InDuDoNet}

\author{Hong Wang\inst{1,2} \and
Yuexiang Li\inst{2}\and
Haimiao Zhang\inst{3}\and
Jiawei Chen\inst{2}\and
Kai Ma\inst{2}\and \\
Deyu Meng\inst{1,4}\textsuperscript{(\Letter)}\and
Yefeng Zheng\inst{2}\textsuperscript{(\Letter)}}
%1{Wang, Hong}
%2{Li, Yuexiang}
%3{Zhang, Haimiao}
%4{Chen, Jiawei}
%5{Ma, Kai}
%6{Meng, Deyu}
%7{Zheng, Yefeng}

\authorrunning{H. Wang et al.}
% First names are abbreviated in the running head.
% If there are more than two authors, 'et al.' is used.
%
%\institute{Princeton University, Princeton NJ 08544, USA \and
%Springer Heidelberg, Tiergartenstr. 17, 69121 Heidelberg, Germany
%\email{lncs@springer.com}\\
%\url{http://www.springer.com/gp/computer-science/lncs} \and
%ABC Institute, Rupert-Karls-University Heidelberg, Heidelberg, Germany\\
%\email{\{abc,lncs\}@uni-heidelberg.de}}
%
\institute{Xi'an Jiaotong University, Xi'an, Shaan'xi, P.R. China\\
\email{hongwang01@stu.xjtu.edu.cn, dymeng@mail.xjtu.edu.cn} \and
Tencent Jarvis Lab, Shenzhen, P.R. China\\
\email{\{vicyxli,kylekma,yefengzheng\}@tencent.com, carvychen@gmail.com} \and
Beijing Information Science and Technology University, Beijing, P.R. China \\
\email{hmzhang@bistu.edu.cn} \and
Macau University of Science and Technology, Taipa, Macau\\
}

%\authorrunning{F. Author et al.}
% First names are abbreviated in the running head.
% If there are more than two authors, 'et al.' is used.
%

%\author{Anonymous}
%\institute{Anonymous Organization\\ \email{**@********.***}}
\maketitle              % typeset the header of the contribution
% \vspace{-5mm}
\begin{abstract}
For the task of metal artifact reduction (MAR), although deep learning (DL)-based methods have achieved promising  performances, most of them suffer from two problems: 1) the CT imaging geometry constraint is not fully embedded into the network during training, leaving room for further performance improvement; 2) the model interpretability is lack of sufficient consideration. Against these issues, we propose a novel interpretable dual domain network, termed as InDuDoNet, which combines the advantages of model-driven and data-driven methodologies. Specifically, we build a joint spatial and Radon domain reconstruction model and utilize the proximal gradient technique to design an iterative algorithm for solving it. The optimization algorithm only consists of simple computational operators, which facilitate us to correspondingly unfold iterative steps into network modules and thus improve the interpretablility of the framework. Extensive experiments on synthesized and clinical data show the superiority of our InDuDoNet. Code is available in \url{https://github.com/hongwang01/InDuDoNet}.%method on the tasks of MAR and downstream multi-class pelvic fracture segmentation.
\end{abstract}

% only containing simple computations  both visually and quantitatively
%
%We are the first to propose a novel model-driven interpretable metal artifact reduction (MAR) network which fully embeds the CT imaging geometry constraint and combines the advantages of traditional prior-based and deep learning-based methodologies. Comprehensive experiments on synthesized and clinic data fully substantiate the superiority and the clinic value of our method both visually and quantitatively for the two tasks of MAR and downstream multi-class pelvic fracture segmentation.
\section{Introduction}
Computed tomography (CT) images reconstructed from X-ray projections play an important role in clinical diagnosis and treatment planning. However, due to the metallic implants within patients, CT images are always adversely affected by undesirable streaking and shading artifacts, which may consequently affect the clinical diagnosis~\cite{de1999metal,park2018ct}. Hence, metal artifact reduction (MAR), as a potential solution, gains increasing attention from the community.
% and then the diagnostic value would be degraded
% an important problem in CT imaging
Various traditional hand-crafted methods~\cite{mehranian2013x, chang2018prior,kalender1987reduction,meyer2010normalized} have been proposed for the MAR task.
% , which can be mainly divided into three categories, \emph{i.e.,} iterative reconstruction, sinogram domain MAR, and image domain MAR. Specifically, iterative algorithms aim at designing some hand-crafted regularizers, such as total variation~\cite{schiffer2014sinogram,zhang2016iterative}, sparsity constraints in the wavelet domain~\cite{zhang2018reweighted}, and formulate them into algorithm optimization. Due to such subjective prior assumptions, these approaches cannot finely capture complicated and diverse metal artifacts in clinic applications. Sinogram domain based methods regard metal-affected regions (i.e., metal trace in sinogram) as missing data and replace them by linear interpolation (LI)~\cite{kalender1987reduction} or forward projection (FP) of a prior image~\cite{meyer2010normalized,wang2013metal}. Yet these surrogate data in the metal trace often do not properly meet the CT imaging geometry constraint and always cause secondary artifacts tangent to the metallic implants in the reconstructed CT images. Image domain methods directly utilize some image processing technologies to suppress the adverse artifacts, which are limited in achieving satisfactory results~\cite{karimi2015metal,soltanian1996ct}.
Driven by the significant success of deep learning (DL) in medical image reconstruction and analysis~\cite{ronneberger2015u,wang2018image,ji2021learning}, researchers began to apply the convolutional neural network (CNN) for MAR in recent years~\cite{zhang2018convolutional,lin2019dudonet,liao2019adn,yu2020deep,lyu2020dudonet++}.
% recent years have also witnessed the rapid progress of convolutional neural network (CNN) for MAR.

Existing deep-learning-based MAR methods can be grouped into three research lines, \emph{i.e.,} sinogram enhancement, image enhancement, and dual enhancement (joint sinogram and image). Concretely, the sinogram-enhancement-based approaches adopt deep networks to directly repair metal-corrupted sinogram~\cite{park2018ct,ghani2019fast,liao2019generative} or utilize the forward projection (FP) of a prior image to correct the sinogram~\cite{gjesteby2017deep, zhang2018convolutional}. For the image enhancement line, researchers exploit the residual learning~\cite{huang2018metal} or adversarial learning~\cite{wang2018conditional,liao2019adn} on CT images only for metal artifact reduction. The dual enhancement of sinogram and image is a recently-emerging direction for MAR. The mutual learning between the sinogram and CT image proposed by recent studies~\cite{lin2019dudonet,yu2020deep,lyu2020dudonet++} significantly boosts the performance of MAR.
% Very recently, there are few works building dual networks to achieve the mutual learning between sinogram and CT image by embedding the differentiate FP layer and filtered back-projection (FBP) layer~\cite{lin2019dudonet,yu2020deep,lyu2020dudonet++}.
% Attributed to the powerful feature representation learning capacity of CNN, these deep-learning-based MAR techniques generally outperform conventional hand-crafted methods. 
Nevertheless, these deep-learning-based MAR techniques share some common drawbacks. The most evident one is that most of them regard MAR as the general image restoration problem and neglect the inherent physical geometry constraints during network training. Yet such constraints are potentially helpful to further boost the performance of MAR.
% across the entire learning process.
Besides, due to the nature of almost black box, the existing approaches relying on the off-the-shelf deep networks are always lack of sufficient model interpretability for the specific MAR task, making them difficult to analyze the intrinsic role of network modules.
% they rely on the off-the-shelf DL toolkits to build different network architectures for this MAR task. It is always difficult to understand and analyze the role of every network unit and thus makes the entire network structure have weak interpretability.

{\color{black}{To alleviate these problems, we propose a novel interpretable dual domain network, termed as InDuDoNet, for the MAR task, which sufficiently embeds the intrinsic imaging geometry model constraints into the process of mutual learning between spatial (image) and Radon (sinogram) domains, and is flexibly integrated with the dual-domain-related prior learning. 
% , which embeds the intrinsic imaging geometry into the process of network training. 
Particularly, we propose a concise dual domain reconstruction model and utilize the proximal gradient technique~\cite{beck2009fast} to design an optimization algorithm. Different from traditional solvers~\cite{zhang2018reweighted} for the model containing heavy operations (\emph{e.g.,} matrix inversion), the proposed algorithm consists of only simple computations (\emph{e.g.,} point-wise multiplication) and thus facilitates us to easily unfold it as a network architecture.}} The specificity of our framework lies in the exact step-by-step corresponding relationship between its modules and the algorithm operations, naturally resulting in its fine physical interpretability.
% , which greatly facilitates a deeper analysis on what happens inside the network training. 
Comprehensive experiments on synthetic and clinical data substantiate the effectiveness of our method. 
% Moreover, we are the first to utilize the downstream multi-class pelvic fracture segmentation task to fairly evaluate the clinic value of our method as compared with other state-of-the-art (SOTA) MAR methods.

% \section{Joint Spatial and Radon Domain Reconstruction model for Metal Artifact Reduction}
\vspace{-0mm}
\section{Method}
In this section, we first theoretically formulate the optimization process for dual domain MAR, and then present the InDuDoNet which is constructed by correspondingly unfolding the optimization process into network modules in details.
% In this section, we introduce the proposed joint spatial and Radon domain reconstruction model in details.

\vspace{1mm}
\noindent{\bf Formulation of Dual Domain Model.}
% \noindent{\bf Model Formulation.}
% In conventional iterative reconstruction, 
Given the observed metal-affected sinogram {{$Y\in \mathbb{R}^{N_{b}\times N_{p}}$}}, where $N_{b}$ and $N_{p}$ are the number of detector bins and projection views, respectively, traditional iterative MAR is formulated as:
\begin{equation}\label{o1}
    \min_{{X}}\left\|(1-Tr)\odot (\mP X-Y)\right\|_{F}^{2} + \lambda g(X),
\end{equation}
where {{$X\in \mathbb{R}^{H\times W}$}} is the clean CT image (\emph{i.e.,} spatial domain); $H$ and $W$ are the height and width of the CT image, respectively; $\mP$ is the Radon transform (\emph{i.e.,} forward projection); $Tr$ is the binary metal trace;  $\odot$ is the point-wise multiplication; $g(\cdot)$ is a regularizer for delivering the prior information of $X$ and $\lambda$ is a trade-off parameter. {\color{black}{For the spatial and Radon domain mutual learning}}, we further execute the joint regularization and transform the problem (\ref{o1}) to:
\begin{equation}\label{o2}
    \min_{{S,X}}\left\|\mP X-S\right\|_{F}^{2} +\alpha \left\|(1-Tr)\odot (S-Y)\right\|_{F}^{2}+ \lambda_{1} g_{1}(S)\!+\lambda_{2} g_{2}(X),
\end{equation}
where $S$ is the clean sinogram (\emph{i.e.,} Radon domain); $\alpha$ is a weight factor balancing the data consistency between spatial and Radon domains; $g_{1}(\cdot)$ and $g_{2}(\cdot)$ are regularizers embedding the priors of the to-be-estimated $S$ and $X$, respectively. %\lambda_{i}$ is the trade-off parameter (i=1,2).

{\color{black}{Clearly, correcting the normalized metal-corrupted sinogram is easier than directly correcting the original metal-affected sinogram, since the former profile is more homogeneous~\cite{meyer2010normalized,zhang2018reweighted}}}. We thus rewrite the sinogram $S$ as:
\begin{equation}\label{YS}
    S=\widetilde{Y}\odot\widetilde{S},
\end{equation}
where $\widetilde{Y}$ is normalization coefficient, usually set as the FP of a prior image $\widetilde{X}$, \emph{i.e.,} $\widetilde{Y} = \mP \widetilde{X}$;\footnote{{\color{black}{We utilize a CNN to flexibly learn $\widetilde{X}$ and $\widetilde{Y}$ from training data as shown in Fig.~\ref{fignet}.}}} $\widetilde{S}$ is the normalized sinogram. By substituting Eq.~(\ref{YS}) into Eq.~(\ref{o2}), we can derive the dual domain reconstruction problem as:
\begin{equation}\label{o3}
\min_{{\widetilde{S}, X}}\left\|\mP X-\widetilde{Y}\odot\widetilde{S}\right\|_{F}^{2} +\alpha \left\|(1-Tr)\!\odot\! (\widetilde{Y}\odot\widetilde{S}\!-\!Y)\right\|_{F}^{2}+ \lambda_{1} g_{1}(\widetilde{S})+\lambda_{2} g_{2}(X).
\end{equation}
As presented in Eq.~(\ref{o3}), our goal is to jointly estimate $\widetilde{S}$ and $X$ from $Y$. {\color{black}{In the traditional prior-based MAR methods, regularizers $g_{1}(\cdot)$ and $g_{2}(\cdot)$ are manually formulated as explicit forms~\cite{zhang2018reweighted}, which cannot always capture complicated  and diverse metal artifacts. Owning to the sufficient and adaptive prior fitting capability of CNN~\cite{wang2020model, xie2020mhf}, we propose to automatically learn the dual-domain-related priors $g_{1}(\cdot)$ and $g_{2}(\cdot)$ from training data using network modules in the following. Similarly, adopting such a data-driven strategy to learn implicit models has been applied in other vision tasks~\cite{wang2021structural, wang2021rain, yue2020variational}}}.

% \vspace{-2mm}
\subsection{Optimization Algorithm}
{\color{black}{Since we want to construct an interpretable deep unfolding network for solving the problem~(\ref{o3}) efficiently, it is critical to build an optimization algorithm with possibly simple operators that can be transformed to network modules easily. Traditional solver~\cite{zhang2018reweighted} for the dual domain model~(\ref{o3}) contains complex operations, \emph{e.g.,} matrix inversion, which are hard for such unfolding transformation. We thus prefer to build a new solution algorithm for problem~(\ref{o3}), which only involves simple computations.}} Particularly, $\widetilde{S}$ and $X$ are alternately updated as:
%To solve the model Eq.~(\ref{o3}), we utilize the proximal gradient technique~\cite{beck2009fast} to alternately update $\widetilde{S}$ and $X$, which is introduced in details in the following:  based on the proximal gradient technique~\cite{beck2009fast} by alternately updating 

\textbf{Updating $\widetilde{S}$}:
The normalized sinogram $\widetilde{S}$ can be updated by solving the quadratic approximation~\cite{beck2009fast} of the problem (\ref{o3}) about $\widetilde{S}$, written as:
\begin{equation}\label{minqs}
  \min_{\widetilde{S}} \frac{1}{2} \left\| \widetilde{S} -\left( \widetilde{S}_{n-1}- \eta_{1}\nabla f\left(\widetilde{S}_{n-1}\right) \right) \right\|_F^2 + \lambda_{1}\eta_{1} g_{1}(\widetilde{S}),
\end{equation}
where {{$\widetilde{S}_{n-1}$}} is the updated result after $(n-1)$ iterations; $\eta_{1}$ is the stepsize parameter; and  {{$f\left(\widetilde{S}_{n-1}\right)\!=\! \left\|\mP X_{n-1}\!-\!\widetilde{Y}\widetilde{S}_{n-1}\right\|_{F}^{2} +\alpha \left\|(1-Tr)(\widetilde{Y}\widetilde{S}_{n-1}\!-\!Y)\right\|_{F}^{2}$}} (note that we omit $\odot$ used in Eq.~(\ref{o3}) for simplicity).
% and the computations about $Tr$ and $\widetilde{Y}$ are all point-wisely. 
For general regularization terms \cite{donoho1995noising}, the solution of Eq. (\ref{minqs}) is:
\begin{equation}\label{sols}
 \widetilde{S}_{n} = \mbox{prox}_{\lambda_{1}\eta_{1}}\left(\widetilde{S}_{n-1} \!-\! \eta_{1}\nabla f\left(\widetilde{S}_{n-1}\right)  \!\right).
\end{equation}
By substituting {\small{$\nabla f\left(\widetilde{S}_{n-1}\right)\!=\!\widetilde{Y}\left(\widetilde{Y}\widetilde{S}_{n-1}-\mP X_{n-1}\right)+\alpha\left(1-Tr\right)\widetilde{Y}\left(\widetilde{Y}\widetilde{S}_{n-1}\!-\!Y\right)$}} into Eq.~(\ref{sols}),
the updating rule of $\widetilde{S}$ is:
\small
\begin{equation}\label{rules}
    \begin{split}
        \widetilde{S}_{n} &= \mbox{prox}_{\lambda_{1}\eta_{1}}\!\left(\widetilde{S}_{n-1} \!-\! \eta_{1}\left(\widetilde{Y}\!\left(\widetilde{Y}\widetilde{S}_{n-1}\!-\!\mP X_{n-1}\right)\!+\!\alpha\left(1-Tr\right)\widetilde{Y}\!\left(\widetilde{Y}\widetilde{S}_{n-1}\!-\!Y\right)  \!\right)\!\right) \\
         &\triangleq \mbox{prox}_{\lambda_{1}\eta_{1}}\!\left(\widehat{S}_{n-1}\right),
    \end{split}
\end{equation}
\normalsize
where $\mbox{prox}_{\lambda_{1}\eta_{1}}(\cdot)$ is the proximal operator related to the regularizer $g_{1}(\cdot)$. Instead of fixed hand-crafted image priors~\cite{zhang2016iterative,zhang2018reweighted}, we adopt a convolutional network module to automatically learn $\mbox{prox}_{\lambda_{1}\eta_{1}}(\cdot)$ from training data (detailed in~{\bf \ref{sec:details}}).

\textbf{Updating $X$}: Also, the image $X$ can be updated by solving the quadratic approximation of Eq.~(\ref{o3}) with respect to $X$:
\small
\begin{equation}\label{minqx}
  \min_{X}\frac{1}{2} \left\| X - \left( X_{n-1}- \eta_{2}\nabla h\left(X_{n-1}\right)  \!\right) \right\|_F^2 + \lambda_{2}\eta_{2} g_{2}(X),
\end{equation}
\normalsize
where {\small{$\nabla h\left(X_{n-1}\right)\!=\!\mP^{T}\left(\mP X_{n-1}-\widetilde{Y}\widetilde{S}_{n}\right)$}}. Thus, the updating formula of $X$ is:
% \small
% \begin{equation}\label{rulex}
%     X_{n} = \mbox{prox}_{\lambda_{2}\eta_{2}}\left(X_{n-1}- \eta_{2}\nabla h\left(X_{n-1}\right)  \right)\triangleq \mbox{prox}_{\lambda_{2}\eta_{2}}\left(\widehat{X}_{n-1}\right),
% \end{equation}
% \normalsize
\small
\begin{equation}\label{rulex}
    X_{n} = \mbox{prox}_{\lambda_{2}\eta_{2}}\left(X_{n-1}- \eta_{2}\mP^{T}\left(\mP X_{n-1}-\widetilde{Y}\widetilde{S}_{n}\right)\right)\triangleq \mbox{prox}_{\lambda_{2}\eta_{2}}\left(\widehat{X}_{n-1}\right),
\end{equation}
\normalsize
where $\mbox{prox}_{\lambda_{2}\eta_{2}}(\cdot)$ is dependent on $g_{2}(\cdot)$. Using the iterative algorithm (Eqs.~(\ref{rules}) and (\ref{rulex})), we can correspondingly construct the deep unfolding network in~{\bf \ref{sec:details}}.

% \vspace{-3mm}
\subsection{Overview of InDuDoNet}\label{sec:details}
\begin{figure*}[t]
  \begin{center}
%   \vspace{-1mm}
     \includegraphics[width=0.94\linewidth]{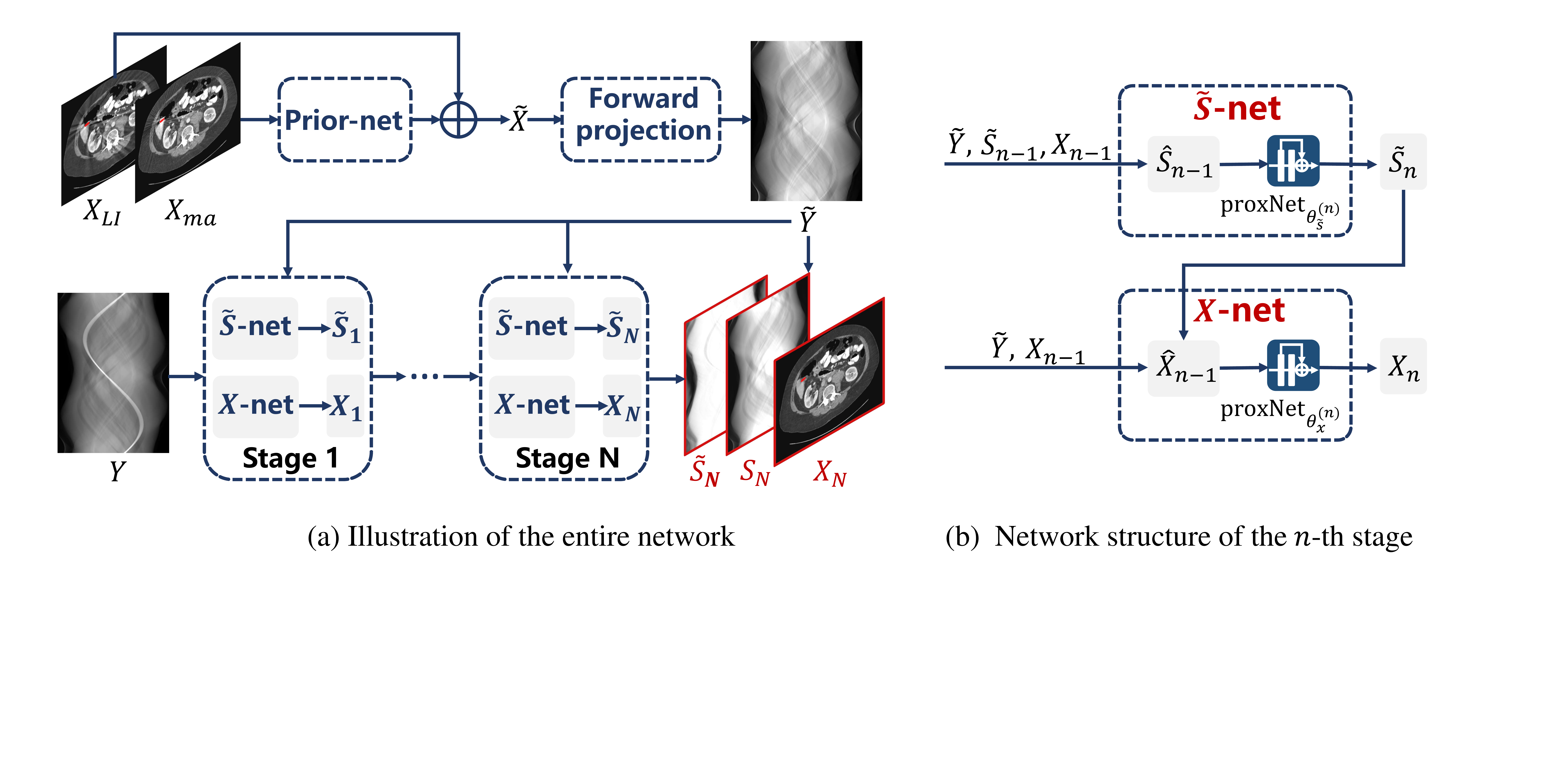}
  \end{center}
  \vspace{-3mm}
     \caption{(a) The proposed network architecture consists of a Prior-net, $N$-stage $\widetilde{S}$-net, and $N$-stage $X$-net. It outputs the normalized sinogram $\widetilde{S}_{N}$, sinogram $S_{N}$, and image $X_{N}$. (b) The detailed structure at the $n$-th stage, in which $\widetilde{S}_{n}$ and $X_{n}$ are successively updated by $\widetilde{S}$-net and $X$-net, respectively, based on the algorithm in Eqs.~(\ref{rules})(\ref{rulex}).}
  \label{fignet}
    \vspace{-1mm}
\end{figure*}

Recent studies~\cite{wang2020model, xie2020mhf} have demonstrated the excellent interpretability of unfolding models. Motivated by these, we propose a deep unfolding framework, namely InDuDoNet, specifically fitting the MAR task. The pipeline of our framework is illustrated in Fig.~\ref{fignet}, which consists of Prior-net, $N$-stage $\widetilde{S}$-net, and $N$-stage $X$-net with parameters $\theta_{prior}$, $\theta_{\widetilde{s}}^{(n)}$, and $\theta_{x}^{(n)}$, respectively. Note that $\widetilde{S}$-net and $X$-net are step-by-step constructed based on the updating rules as expressed in Eqs.~(\ref{rules}) and (\ref{rulex}), which results in a specific physical interpretability of our framework. All the parameters including $\theta_{prior}$, $\{\theta_{\widetilde{s}}^{(n)},\theta_x^{(n)}\}_{n=1}^{N}$, $\eta_{1}$, $\eta_{2}$, and $\alpha$ can be automatically learned from the training data in an end-to-end manner.
% Motivated by the recent studies~\cite{yang2017admm,yang2018proximal}, we propose a deep unfolding framework, namely InDuDoNet, for the MAR task.
\noindent\textbf{Prior-net.} Prior-net in Fig.~\ref{fignet} is utilized to learn $\widetilde{Y}$ from the concatenation of metal-affected image $X_{ma}$ and linear interpolation (LI) corrected image $X_{LI}$~\cite{kalender1987reduction}. Our Prior-net has a similar U-shape architecture~\cite{ronneberger2015u} to the PriorNet in~\cite{yu2020deep}.
%proposed by Yu \emph{et al.}
% U-Net~\cite{ronneberger2015u} is adopted as the backbone for Prior-net.
% we simply utilize the U-Net~\cite{ronneberger2015u} to learn $\widetilde{X}$ by taking the concatenation of metal-affected image $X_{ma}$ and LI corrected image $X_{LI}$ as input, where $X_{ma}$ and $X_{LI}$ are reconstructed from original metal-corrupted sinogram $Y$ and the linear interpolated sinogram $Y_{LI}$~\cite{kalender1987reduction}. Here the adopted U-Net has the similar structure as PriorNet in~\cite{yu2020deep} with the depth of 4,  but we halve the channel number for fewer network parameters.

\noindent{\bf $\widetilde{S}$-net and $X$-net.} With $\widetilde{Y}$ generated by Prior-net, the framework reconstructs the artifact-reduced sinogram $\widetilde{S}$ and the CT image $X$ via sequential updates of $\widetilde{S}$-net and $X$-net. As shown in Fig.~\ref{fignet}(a), $N$ stages are involved in our framework, which correspond to $N$ iterations of the algorithm for solving~(\ref{o3}). Each stage shown in Fig.~\ref{fignet}(b)
%, together with the learning of $\widetilde{Y}$.
% We now introduce the network architecture for each stage. 
% One can refer to Fig.~\ref{fignet}(b) for better understanding. 
% As shown in Fig.~\ref{fignet} (b), 
% Each stage yields the sequential updates of $\widetilde{S}$ and $X$ generated by $\widetilde{S}$-net and $X$-net, respectively. 
is constructed by unfolding the updating rules Eqs.~(\ref{rules}) and (\ref{rulex}), respectively.
Particularly, for the $n$-th stage, $\widehat{S}_{n-1}$ is firstly computed based on Eq.~(\ref{rules}) and then fed to a deep network {\small{$\text{proxNet}_{\theta_{\widetilde{s}}^{(n)}}(\cdot)$}} to execute the operator $\mbox{prox}_{\lambda_{1}\eta_{1}}(\cdot)$. Then, we obtain the updated normalized sinogram:
{\small {$\widetilde{S}_{n} = \text{proxNet}_{\theta_{\widetilde{s}}^{(n)}}\left(\widehat{S}_{n-1}\right)$}}. Similar operation is taken to process {\small{$\widehat{X}_{n-1}$}} computed based on Eq.~(\ref{rulex}) and the updated artifact-reduced image is: {\small{$X_{n} = \text{proxNet}_{\theta_{x}^{(n)}}\left(\widehat{X}_{n-1}\right)$}}. {\small{$\text{proxNet}_{\theta_{\widetilde{s}}^{(n)}}(\cdot)$}} and {\small{$\text{proxNet}_{\theta_{x}^{(n)}}(\cdot)$}} have the same structure---four [{\small{\emph{Conv+BN+ReLU\\+Conv+BN+Skip Connection}}}] residual blocks~\cite{he2016deep}. After $N$ stages of optimization, the framework can well reconstruct the normalized sinogram $\widetilde{S}_{N}$, and therefore yield the final sinogram {\small{${S}_{N}$}} by {\small{$\widetilde{Y}\odot\widetilde{S}_{N}$}} (refer to Eq.~(\ref{YS})), and the CT image {\small{$X_{N}$}}.
%  using $\widetilde{X}$ learnt from the Prior-net.% with parameter $\theta_{prior}$.

{\color{black}{\noindent\emph{\textbf{Remark:}} Our network is expected to possess both the advantages of the model-driven and data-driven methodologies. Particularly, compared with traditional prior-based methods, our network can flexibly learn sinogram-related and image-related priors through {\small{$\text{proxNet}_{\theta_{\widetilde{s}}^{(n)}}(\cdot)$}} and {\small{$\text{proxNet}_{\theta_{x}^{(n)}}(\cdot)$}} from training data. Compared with deep MAR methods, our framework incorporates both CT imaging constraints and dual-domain-related priors into the network architecture.}}

% first computed based on Eq.~(\ref{rulex}) and then sent into a deep network $\text{proxNet}_{\theta_{x}^{(n)}}(\cdot)$ with parameter $\theta_{x}^{(n)}$ to perform the mapping $\mbox{prox}_{\lambda_{2}\eta_{2}}(\cdot)$. Then, the updated artifact-reduced image is: $X_{n} = \text{proxNet}_{\theta_{x}^{(n)}}\left(\widehat{X}_{n-1}\right)$.
% {\itshape $\widetilde{S}$-net and $X$-net.}
% with the channel number as 32 and convolutional kernel size as $3\times3$.

\noindent\textbf{Training Loss.} We adopt the mean square error (MSE) for the extracted sinogram $\widetilde{Y}\odot\widetilde{S}_{n}$ and image $X_n$  at every stage as the training objective function:
\small
\begin{equation}\label{Loss}
%\vspace{-1mm}
  \mathcal{L} = \sum_{n=0}^{N}\beta_{n}\left\|X_n-X_{gt} \right\|_F^2\odot(1-M)+\gamma\left(\sum_{n=1}^{N}\beta_{n}\left\| \widetilde{Y}\odot\widetilde{S}_n - Y_{gt}\right\|_F^2\right),
%\vspace{-1mm}
\end{equation}
\normalsize
where $X_{gt}$ and $Y_{gt}$ are ground truth image and metal-free sinogram, respectively. We simply set $\beta_{N}=1$ to make the outputs at the final stage play a dominant role, and $\beta_{n}=0.1$ ($n=0,\cdots, N-1$) to supervise each middle stage. $\gamma$ is a hyperparamter to balance the weight of different loss items and we empirically set it as 0.1. We initialize $X_{0}$ by passing $X_{LI}$ through a proximal network {\small{$\text{proxNet}_{\theta_{x}^{(0)}}$}}.
%\textbf{Differences from Existing Iterative Methods}
% \vspace{-3mm}
\section{Experimental Results}
% \vspace{-2mm}
% \subsection{Dataset and Implementation Details}
\textbf{Synthesized Data.} Following the simulation protocol in~\cite{yu2020deep}, we randomly select a subset from the DeepLesion~\cite{yan2018deep} to synthesize metal artifact data. The metal masks are from~\cite{zhang2018convolutional}, which contain 100 metallic implants with different shapes and sizes. We choose 1,000 images and 90 metal masks to synthesize the training samples, and pair the additional 200 CT images from 12 patients with the remaining 10 metal masks to generate 2,000 images for testing. The sizes of the 10 metallic implants for test data are: [2061, 890, 881, 451, 254, 124, 118, 112, 53, 35] in pixels. Consistent to~\cite{lin2019dudonet,lyu2020dudonet++}, we simply put the adjacent sizes into one group when reporting MAR performance. We adopt the procedures widely used by existing studies~\cite{zhang2018convolutional,liao2019adn,lin2019dudonet,yu2020deep,lyu2020dudonet++} to simulate $Y$ and $X_{ma}$. 
% Various effects are considered, including polychromatic X-ray, partial volume effect, beam hardening, and Possion noise. 
All the CT images are resized to $416\times 416$ pixels and 640 projection views are uniformly spaced in 360 degrees. The resulting sinograms are of the size $N_{b}\times N_{p}$ as $641\times 640$.
% between 0-360 degrees. 
%\footnote{The simulation code is borrowed from the authors~\cite{yu2020deep}}.

\vspace{1mm}
\noindent\textbf{Clinical Data.} We further assess the feasibility of the proposed InDuDoNet on a clinical dataset, called CLINIC-metal~\cite{liu2020deep}, for pelvic fracture segmentation. The dataset includes 14 testing volumes labeled with multi-bone, \emph{i.e.,} sacrum, left hip, right hip, and lumbar spine. The clinical images are resized and processed using the same protocol to the synthesized data. Similar to~\cite{liao2019adn,yu2020deep}, the clinical metal masks are segmented with a thresholding (2,500 HU).

%and then we can subsequently obtain the $Tr$ via forward projection (FP), $S_{ma}$ via filtered backprojection (FBP) operation provided in ODL library\footnote{\url{https://github.com/odlgroup/odl}}, and $S_{LI}$ via linear interpolation~\cite{kalender1987reduction}.
\vspace{1mm}
\noindent\textbf{Evaluation Metrics.} The peak signal-to-noise ratio (PSNR) and structured similarity index (SSIM) with the code from~\cite{zhang2018convolutional} are adopted to evaluate the performance of MAR. Since we perform the downstream multi-class segmentation on the CLINIC-metal dataset to assess the improvement generated by different MAR approaches to clinical applications, the Dice coefficient (DC) is adopted as the metric for the evaluation of segmentation performance.

% For DeepLesion, we adopt the peak signal-to-noise ratio (PSNR) and structured similarity index (SSIM) with the code from~\cite{zhang2018convolutional}  for quantitative evaluation. For CLINIC-metal, we first execute the task of MAR and only provide the visual comparison since the clean images are not available. Then we perform the downstream multi-class segmentation on the artifact-removed results and do quantitative evaluation with the metric as Dice coefficient (DC).
{\color{black}{\noindent\textbf{Training Details.} Based on a NVIDIA Tesla V100-SMX2 GPU, we implement our network with PyTorch~\cite{paszke2017automatic} and differential operations $\mP$ and $\mP^T$ in ODL library.\footnote{\url{https://github.com/odlgroup/odl}.}}} We adopt the Adam optimizer with ($\beta_{1}$, $\beta_{2}$)=(0.5, 0.999). The initial learning rate is $2\times10^{-4}$ and divided by 2 every 40 epochs. The total epoch is 100 with a batch size of 1. Similar to~\cite{yu2020deep}, in each training iteration, we randomly select an image and a metal mask to synthesize a metal-affected sample.

% {\color{black}{\noindent\textbf{Training Details.} We implement our network based on PyTorch~\cite{paszke2017automatic} in an end-to-end manner with differential operations $\mP$ and $\mP^T$ in ODL library\footnote{\url{https://github.com/odlgroup/odl}}. We adopt the Adam optimizer with ($\beta_{1}$,$\beta_{2}$)=(0.5,0.999). The initial learning rate is $2\times10^{-4}$ and divided by 2 every 40 epochs. The total epoch is 100 with a batch size of 2. Similar to~\cite{yu2020deep}, in each training iteration, we randomly select one image from the pool of 1000 images and one metal mask from the pool of 90 masks to synthesize metal artifacts. Different CT images are formed as one batch data.}}
\begin{table}[t]
\centering
\caption{Effect of the total stage number $N$ on the performance of the proposed InDuDoNet on synthesized data with PSNR (dB) and SSIM.}%\vspace{2mm}  % Bold and underline indicate top $1^{\text{st}}$ and $2^{\text{nd}}$ rank, respectively.
% $^{\text{*}}$ means we adopt the pre-trained model released by the authors \cite{wei2019semi}.
%\begin{tabular}{@{}c|c@{}c|c|c|c|c|c|c@{}}
\tiny
\setlength{\tabcolsep}{6.5pt}
\begin{tabular}{l|c|c|c|c|c|c}
\Xhline{0.6pt}
   $N$  & \multicolumn{5}{c|}{ Large Metal \quad \quad   \quad\quad  $\longrightarrow$    \quad   \quad\quad \quad         Small Metal}                & Average     \\
\Xhline{0.6pt}
$N$=0              &28.91/0.9280              &30.42/0.9400              &34.45/0.9599               &36.72/0.9653              &37.18/0.9673              &33.54/0.9521              \\
$N$=1              &34.10/0.9552              &35.91/0.9726              &38.48/0.9820              &39.94/0.9829              &40.39/0.9856              &37.76/0.9757              \\
$N$=3            &33.46/0.9564  &37.14/0.9769  &40.33/0.9868              &42.55/0.9896              &42.68/0.9908             &39.23/0.9801              \\
$N$=6          & 34.59/0.9764 & 38.95/0.9890 & 42.28/0.9941  & 44.09/0.9945 & 45.09/0.9953 & 41.00/\underline{0.9899} \\
$N$=10          &36.74/0.9801 & 39.32/0.9896 & 41.86/0.9931 & 44.47/0.9942 &45.01/0.9948 &\underline{41.48}/\textbf{0.9904}  \\
%$N$=12          &36.06/0.9709 & 39.93/0.9896 &43.85/0.9955 & 45.58/0.9960 & 45.86/0.9967 &\textbf{42.25}/0.9897  \\
$N$=12          &36.52/0.9709 & 40.01/0.9896 &42.66/0.9955 & 44.17/0.9960 & 44.84/0.9967 &\textbf{41.64}/0.9897  \\
\Xhline{0.6pt}
\end{tabular}
\label{tabN}
\vspace{-3mm}
\end{table}
% \begin{table}[t]
% \centering
% \caption{Effect of stage number $N$ on the performance of the proposed network.}  % Bold and underline indicate top $1^{\text{st}}$ and $2^{\text{nd}}$ rank, respectively.
% % $^{\text{*}}$ means we adopt the pre-trained model released by the authors \cite{wei2019semi}.
% %\begin{tabular}{@{}c|c@{}c|c|c|c|c|c|c@{}}
% \tiny
% \setlength{\tabcolsep}{4pt}
% \begin{tabular}{l|c|c}
% \Xhline{0.9pt}
% Methods    &Parameter\#   & Test Time(Seconds) \\
% \Xhline{0.9pt}
% DSCMAR                &25,834,251      &0.3638       \\
% Ours($N$=6)           &4,541,060         &0.3526       \\
% Ours($N$=10)          &5,174,936       &0.5116       \\
% \Xhline{0.9pt}
% \end{tabular}
% \label{tabN}
% \end{table}

%\begin{table}[t]
%\centering
%\caption{Effect of Resblocks number $T$, involved in the ResNet as shown in Fig.~\ref{mbres}, on the performance of the proposed network.}
%\footnotesize
%\setlength{\tabcolsep}{8.8pt}
%\begin{tabular}{c|c|c|c|c|c}
%  % after \\: \hline or \cline{col1-col2} \cline{col3-col4} ...
%\hline
%$T$ & $T$=1& $T$=2 & $T$=3 & $T$=4 & $T$=5\\
%\hline
%PSNR & 39.04 &39.52 &39.80 &40.00 &39.98\\
%\hline
%SSIM  &0.9833 &0.9848 &0.9856 &0.9860 &0.9859\\
%\hline
%\end{tabular}
%\label{tabT}
%\end{table}
%\vspace{-3mm}
\subsection{Ablation Study}
% \vspace{-2mm}
Table~\ref{tabN} lists the performance of our framework under different stage number $N$.
% reports the effect of stage number $N$ on the MAR performance of our method. 
The $N=0$ entry means that the initialization $X_0$ is directly regarded as the reconstruction result. Taking $N=0$ as the baseline, we can find that with only one stage ($N=1$), the MAR performance yielded by our proposed InDuDoNet is already evidently improved, which validates the essential role of the mutual learning between $\widetilde{S}$-net and $X$-net. When $N=12$, the SSIM is slightly lower than that of $N=10$. The underlying reason is that the more stages cause a deeper network and may suffer from gradient vanishing. Hence, {\color{black}{for better performance and fewer network parameters}}, we choose $N=10$ in all our experiments.\footnote{More analysis on network parameter and testing time are in {\itshape supplementary material}.}

% our method achieves significant artifact removal performance, 
\begin{figure*}[t]
  \begin{center}
%   \vspace{-1mm}
     \includegraphics[width=0.9\linewidth]{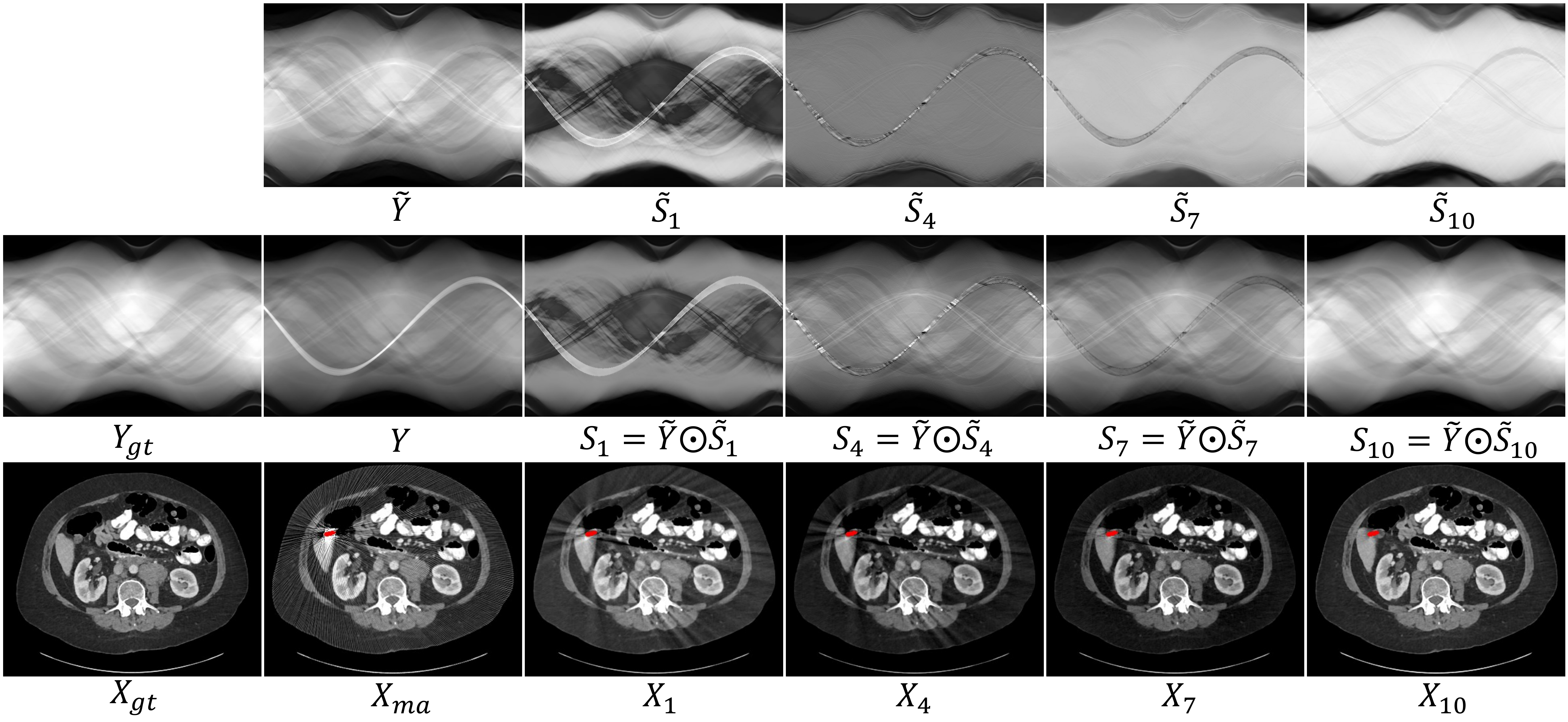}
  \end{center}
  \vspace{-6mm}
     \caption{The recovery normalization coefficient $\widetilde{Y}$, normalized sinogram $\widetilde{S}_{n}$, sinogram $S_{n}$, and image $X_{n}$ at different stages ($N$=10). The red pixels stand for metallic implant.}
  \label{figmodelver}
    \vspace{1mm}
\end{figure*}
\begin{table}[t]
\centering
\caption{PSNR (dB) and SSIM of different MAR methods on synthesized data.}%\vspace{2mm}
% $^{\text{*}}$ means we adopt the pre-trained model released by the authors \cite{wei2019semi}.
%\begin{tabular}{@{}c|c@{}c|c|c|c|c|c|c@{}}
\tiny
\setlength{\tabcolsep}{1.5pt}
\begin{tabular}{l|c|c|c|c|c|c}
\Xhline{0.6pt}
Methods    & \multicolumn{5}{c|}{ Large Metal \quad \quad   \quad\quad  $\longrightarrow$    \quad   \quad\quad \quad         Small Metal}                & Average      \\
\Xhline{0.6pt}
Input             &24.12/0.6761              &26.13/0.7471              &27.75/0.7659               &28.53/0.7964              &28.78/0.8076              &27.06/0.7586              \\
LI~\cite{kalender1987reduction}              &27.21/0.8920              &28.31/0.9185              &29.86/0.9464              &30.40/0.9555              &30.57/0.9608              &29.27/0.9347\\
NMAR~\cite{meyer2010normalized}              &27.66/0.9114              &28.81/0.9373              &29.69/0.9465              &30.44/0.9591              &30.79/0.9669              &29.48/0.9442              \\

CNNMAR~\cite{zhang2018convolutional}            &28.92/0.9433  &29.89/0.9588  & 30.84/0.9706             &31.11/0.9743              &31.14/0.9752              &30.38/0.9644              \\
DuDoNet~\cite{lin2019dudonet}           & 29.87/0.9723 & 30.60/0.9786 & 31.46/0.9839  & 31.85/0.9858 & 31.91/0.9862 & 31.14/0.9814 \\
DSCMAR~\cite{yu2020deep}           & 34.04/0.9343 & 33.10/0.9362 & 33.37/0.9384  & 32.75/0.9393 & 32.77/0.9395 & 33.21/0.9375 \\
DuDoNet++~\cite{lyu2020dudonet++}         & \underline{36.17}/\underline{0.9784} & \underline{38.34}/\underline{0.9891} & \underline{40.32}/\underline{0.9913}  & \underline{41.56}/\underline{0.9919} & \underline{42.08}/\underline{0.9921} & \underline{39.69}/\underline{0.9886} \\
InDuDoNet (Ours) & \textbf{36.74/0.9801} & \textbf{39.32}/\textbf{0.9896} & \textbf{41.86}/\textbf{0.9931} & \textbf{44.47}/\textbf{0.9942} & \textbf{45.01}/\textbf{0.9948}& \textbf{41.48}/\textbf{0.9904} \\ \Xhline{0.6pt}
\end{tabular}
\vspace{-1mm}
\label{tabsyn}
\end{table}
\begin{figure*}[t]
  \begin{center}
  %\vspace{-3mm}
     \includegraphics[width=0.95\linewidth]{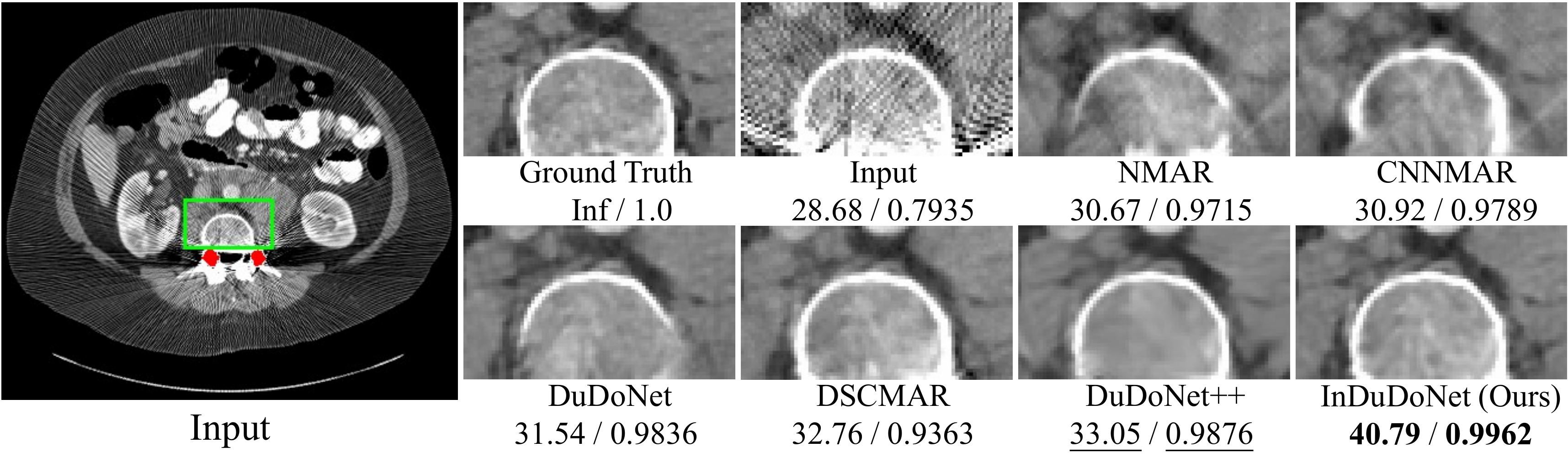}
  \end{center}
  \vspace{-6mm}
     \caption{Comparison for medium metallic implants. PSNR (dB)/SSIM below is for reference. The display window is [-175, 275] HU. The red pixels stand for metallic implants.}
  \label{figsyn}
    \vspace{1mm}
\end{figure*}
%Bold and underline indicate top $1^{\text{st}}$ and $2^{\text{nd}}$ rank, respectively.
% \vspace{-4mm}\subsection{Model Verification}\vspace{-2mm} extracted normalization coefficient $\widetilde{Y}$ and 
\noindent{\bf Model Verification.}
We conduct a model verification experiment to present the mechanism underlying the network modules ($\widetilde{S}$-net and $X$-net). The evaluation results are shown in Fig.~\ref{figmodelver}. The normalized sinogram $\widetilde{S}_{n}$, sinogram $S_{n}$, and CT image $X_{n}$ generated at different stages ($n=1,4,7,10$) are presented on the first, second and third rows, respectively. It can be observed that the metal trace region in $\widetilde{S}_n$ is gradually flattened as $n$ increases, which correspondingly ameliorates the sinogram $S_n$. Thus, the metal artifacts contained in the CT image $X_n$ are gradually removed. The results verify the design of our interpretable iterative learning framework---the mutual promotion of $\widetilde{S}$-net and $X$-net enables the proposed InDuDoNet to achieve MAR along the direction specified by Eq.~(\ref{o3}).
% $1^{\text{st}}$, $2^{\text{nd}}$ and $3^{\text{rd}}$ rows, respectively.

% We utilize synthesized data to execute model verification for presenting the working mechanism inside the network modules. Fig.~\ref{figmodelver} presents the extracted normalization coefficient $\widetilde{Y}$ and normalized sinogram $\widetilde{S}_{n}$($1^{\text{st}}$ row), sinogram $S_{n}$ ($2^{\text{nd}}$ row), and image $X_{n}$($3^{\text{rd}}$ row) at different stages ($n=1,4,7,10$). As shown, with the increase of $n$, $\widetilde{S}_n$ is becoming flatter near the metal trace region and contains more useful information. Correspondingly, the sinogram $S_n$ is gradually ameliorated and $X_n$ covers fewer metal artifacts and more image details. This interpretable learning process should be attributed to the proper design of the joint spatial and Radon domain reconstruction model. The mutual promotion of $\widetilde{S}$-net and $X$-net indeed enables the entire network evolve in a right direction.

\begin{table}[t]
   \vspace{-5mm}
\centering
\caption{The Dice coefficient (DC) results on the downstream segmentation task.}%\vspace{2mm}
\scriptsize
\setlength{\tabcolsep}{1.6pt}
\begin{tabular}{c|ccccccc|c}
\Xhline{0.6pt}
%\hline
Bone       & Input  & LI     & NMAR  & CNNMAR & DuDoNet & DSCMAR                      & DuDoNet++    & InDuDoNet\\
\Xhline{0.6pt} 
Sacrum &0.9247 & 0.9086   & 0.9151 & 0.9244                     & 0.9326  & 0.9252  & \textbf{0.9350}  &\underline{0.9348}\\
Left hip     & 0.9543 & 0.9391 & 0.9427                   & 0.9485                     & 0.9611  & 0.9533  & \underline{0.9617}     &\textbf{0.9630}   \\
Right hip    & 0.8747 & 0.9123 & 0.9168                  & 0.9250                     & \underline{0.9389} & 0.9322  & 0.9379     &\textbf{0.9421}   \\
Lumbar spine & 0.9443 & 0.9453 & 0.9464                   & 0.9489                     & 0.9551  & 0.9475  &\textbf{{0.9564}}     &\underline{0.9562} \\
\Xhline{0.6pt}
Average DC & 0.9245 & 0.9263 & 0.9303                   & 0.9367                     & 0.9469  & 0.9396  & \underline{0.9478}  &\textbf{0.9490}  \\
\Xhline{0.6pt}
\end{tabular}
\label{tabclinic}
    \vspace{-3mm}
\end{table}

% \vspace{-3mm}
\subsection{Performance Evaluation}
% \vspace{-2mm}
{\bf Synthesized Data.} We compare the proposed InDuDoNet with current state-of-the-art (SOTA) MAR approaches, including traditional LI~\cite{kalender1987reduction} and NMAR~\cite{meyer2010normalized}, DL-based CNNMAR~\cite{zhang2018convolutional}, DuDoNet~\cite{lin2019dudonet}, DSCMAR~\cite{yu2020deep}, and DuDoNet++~\cite{lyu2020dudonet++}. For LI, NMAR, and CNNMAR, we directly use the released code and model. We re-implement DuDoNet, DSCMAR, and DuDoNet++, since there is no official code.
% \noindent\textbf{Quantitative Comparison.} \footnote{The network model is borrowed from the authors~\cite{yu2020deep}.}
Table~\ref{tabsyn} reports the quantitative comparison. We can observe that most of DL-based methods consistently outperform the conventional LI and NMAR, showing the superiority of data-driven deep CNN for MAR. The dual enhancement approaches (\emph{i.e.,} DuDoNet, DSCMAR, and DuDoNet++) achieve higher PSNR than the sinogram-enhancement-only CNNMAR. Compared to DuDoNet, DSCMAR, and DuDoNet++, our dual-domain method explicitly embeds the physical CT imaging geometry constraints into the mutual learning between spatial and Radon domains, \emph{i.e.,} jointly regularizing the sinogram and CT image recovered at each stage. Hence, our method achieves the highest PSNRs and SSIMs for all metal sizes as listed. The visual comparisons are shown in Fig.~\ref{figsyn}.\textsuperscript{\ref{exp}}

\noindent{\bf Clinical Data.}
% \vspace{-3mm}\subsection{Experiments on Clinical Data}\vspace{-2mm} MAR as well as 
We further evaluate all MAR methods on clinical downstream pelvic fracture segmentation task using the CLINIC-metal dataset. A U-Net is firstly trained using the clinical metal-free dataset (CLINIC~\cite{liu2020deep}) and then tested on the {\color{black}{metal-artifact-reduced CLINIC-metal CT images generated by different MAR approaches.}}
% Specifically, we utilize CLINIC-metal to qualitatively compare the artifact-removed results. Then we adopt UNet to train an effective pelvic fracture segmentation baseline\footnote{With our baseline, the segmentation accuracy of the original input in Table~\ref{tabclinic} later is comparable to that in~\cite{liu2020deep}, which confirms the rationality of the designed baseline.} based on the clinical metal-free dataset, i.e., CLINIC~\cite{liu2020deep}, and quantitatively evaluate the clinical feasibility of all comparison methods based on segmentation accuracy.
% \noindent\textbf{Quantitative Comparison.} 
The segmentation accuracy achieved by the metal-free-trained U-Net is reported in Table~\ref{tabclinic}.
% reports the segmentation accuracy of the artifact-reduced images by all comparison methods on four-class bone structures, including scrum, left hip, right hip, and lumbar spine. 
We can observe that in average, our method finely outperforms other SOTA approaches. This comparison fairly demonstrates that {\color{black}{our network generalizes well for clinical images with unknown metal materials and geometries and}}  is potentially useful for clinical applications.\footnote{More comparisons of MAR and bone segmentation are in {\itshape supplementary material}\label{exp}.}

\vspace{-1mm}
\section{Conclusion}
% In this paper, we proposed a joint spatial and Radon domain reconstruction model, namely InDuDoNet, for this metal artifact reduction (MAR) task and design an iterative optimization algorithm, which improves the interpretability of our framework. Extensive experiments were conducted on not only the synthesized, but also the clinical data. The experimental results demonstrated the effectiveness of our dual-domain MAR approach---a new state-of-the-art was achieved.
{\color{black}{In this paper, we have proposed a joint spatial and Radon domain reconstruction model for the metal artifact reduction (MAR) task and constructed an interpretable network architecture, namely InDuDoNet, by unfolding an iterative optimization algorithm with only simple computations involved.}} Extensive experiments were conducted on synthesized and clinical data. The experimental results demonstrated the effectiveness of our dual-domain MAR approach as well as its superior interpretability beyond current SOTA deep MAR networks.

\vspace{-2mm}
\subsubsection{Acknowledgements.} This research was supported by National Key R\&D Program of China (2020YFA0713900), the Macao Science and Technology Development Fund under Grant 061/2020/A2, Key-Area Research and Development Program of Guangdong Province, China (No. 2018B010111001),  the Scientific and Technical Innovation 2030-``New Generation Artificial Intelligence''  Project (No. 2020AAA0104100), the China NSFC projects (62076196, 11690011,61721002, U1811461).

% By correspondingly unfolding every step involved in this algorithm into each network module, we have easily constructed the entire network architecture with fine interpretability. This has been visually verified by model visualization for easy understanding the working mechanism of our network. Based on synthesized and clinic datasets, comprehensive experiments on MAR task as well as downstream multi-class pelvic fracture segmentation task, have shown great superiority of our method as compared with current SOTA MAR approaches both visually and quantitatively.
% ---- Bibliography ----
%
% BibTeX users should specify bibliography style 'splncs04'.
% References will then be sorted and formatted in the correct style.

\bibliographystyle{splncs04}
\bibliography{egbib}
%\newpage

\section*{Supplementary Material}
\begin{table}[h]
	\vspace{-2mm}
	\centering
	\caption{Comparison of network parameter and testing time. The average inference time is computed on 2000 images with size $416\times416$ based on a NVIDIA Tesla V100-SMX2 GPU.}%\vspace{2mm}  % Bold and underline indicate top $1^{\text{st}}$ and $2^{\text{nd}}$ rank, respectively.
	% $^{\text{*}}$ means we adopt the pre-trained model released by the authors \cite{wei2019semi}.
	%\begin{tabular}{@{}c|c@{}c|c|c|c|c|c|c@{}}
	\scriptsize
	\setlength{\tabcolsep}{3.5pt}
	\begin{tabular}{l|c|c|c}
		\Xhline{0.6pt}
		Methods    & Network Parameters\#   & Test Time (Seconds)  & Average PSNR (dB)/SSIM\\
		\Xhline{0.6pt}
		DSCMAR                &25,834,251      &0.3638    &33.21/0.9375   \\
		DuDoNet++  &25,983,627  &0.8062   &39.69/0.9886 \\
		InDuDoNet ($N$=6)           &4,541,060         &0.3526   & \underline{41.00}/\underline{0.9892}  \\
		InDuDoNet ($N$=10)          &5,174,936       &0.5116     & \textbf{41.48}/\textbf{0.9904} \\
		\Xhline{0.6pt}
	\end{tabular}
	\vspace{-0mm}
	\label{tabNsupp}
\end{table}

\begin{figure*}[h]
	\begin{center}
		\vspace{-5mm}
		\includegraphics[width=1\linewidth]{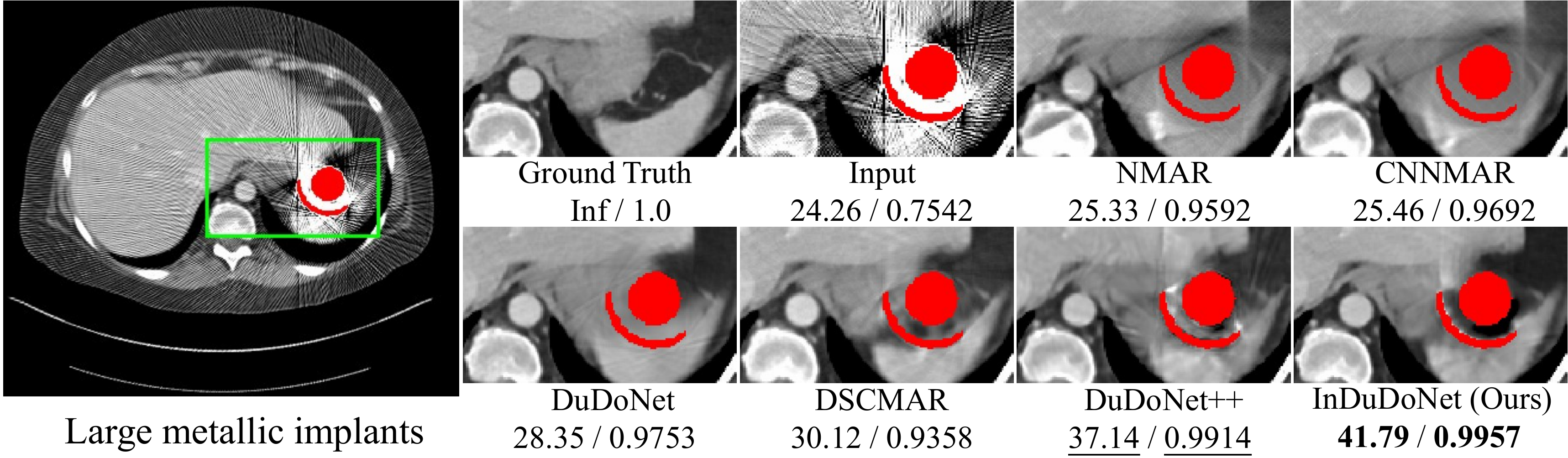}
	\end{center}
	\vspace{-5mm}
	\caption{From the regions around the red metallic implant, we can observe that our method preserves more tissue structures and recovers anatomically more faithful image.}
	\label{figsynsupp}
	\vspace{-1mm}
\end{figure*}
\begin{figure*}[h]
	\begin{center}
		\vspace{-3mm}
		\includegraphics[width=1\linewidth]{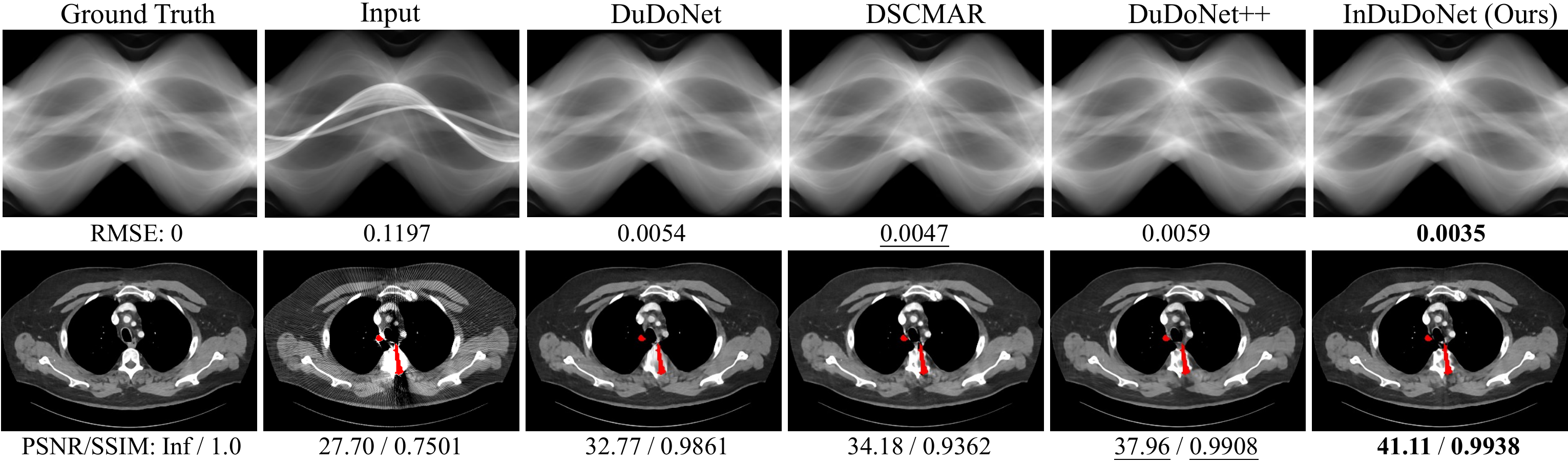}
	\end{center}
	\vspace{-4mm}
	\caption{Comparison of different dual-domain-enhancement-based MAR methods. RMSE for sinogram and PSNR (dB)/SSIM for image below are for reference.}
	\label{figsynsupp}
	\vspace{-1mm}
\end{figure*}
\begin{figure*}[h]
	\begin{center}
		\vspace{-6mm}
		\includegraphics[width=1\linewidth]{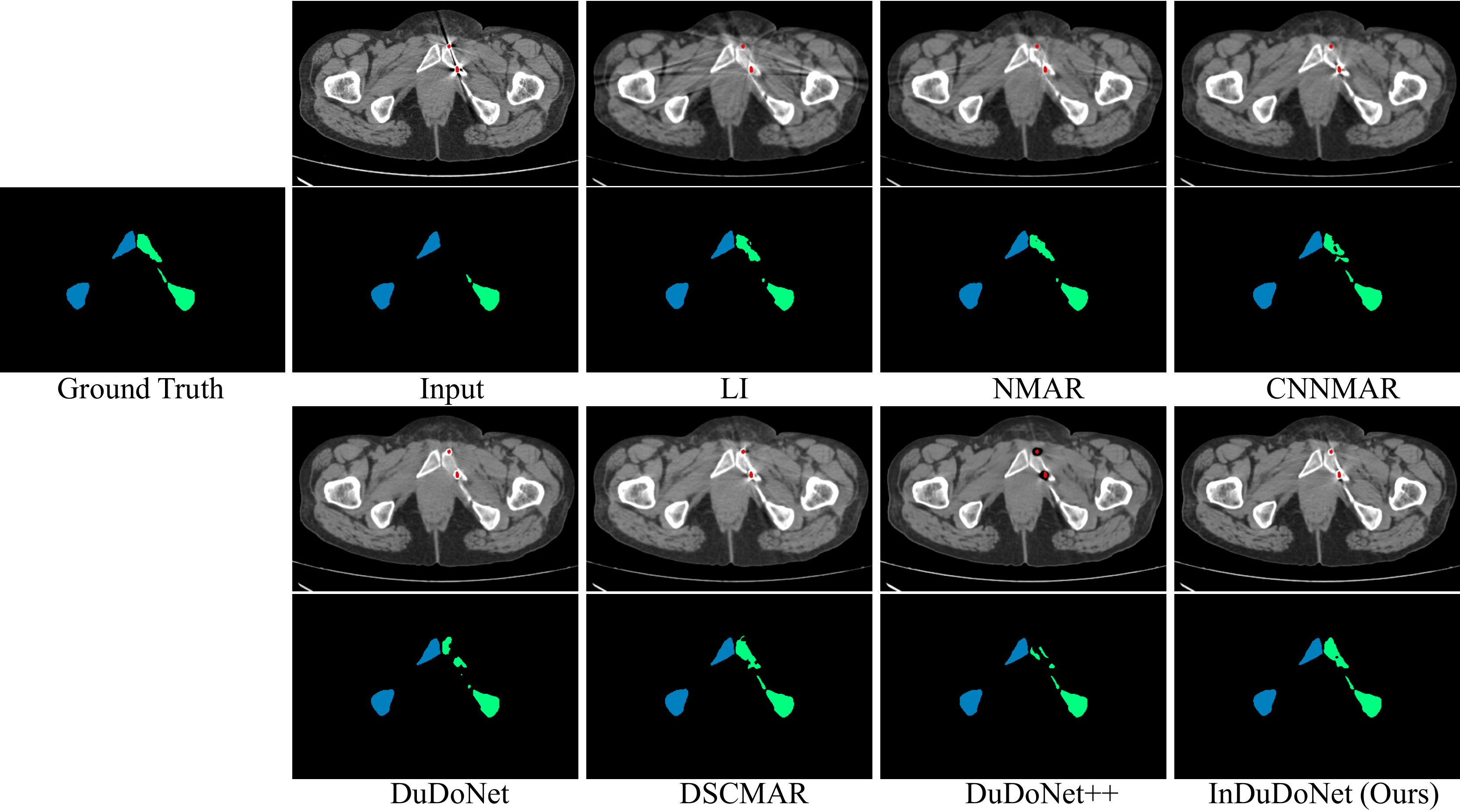}
	\end{center}
	\vspace{-3mm}
	\caption{$1^{\text{st}}$ row is the generalization result on a CLINIC-metal CT image and $2^{\text{nd}}$ row is the segmentation result.  The clinical metal masks are colored in red for better visualization. As observed, comparison methods with the inaccurate recovery of right hip, such as residual metal artifacts or missing structure, have the unsatisfactory segmentation results. However, the restored image by our method shows more credible contents with more details, which effectively promotes the segmentation performance. }
	\label{figclinicsupp}
	\vspace{-0mm}
\end{figure*}

\end{document}